\documentclass{PoS}
\usepackage{aas_macros}
\newcommand{\lsi}{LS I +61$^{\circ}$303 }
\newcommand{\cms}{$\mathrm{cm^{2} s^{-1}}$}

\title{Decadal VERITAS Observation of LS I +61$^{\circ}$303 : Detection of TeV emission around the entire orbit}

\ShortTitle{Decadal VERITAS Observation of \lsi}

\author{\speaker{Payel Kar} {for the VERITAS Collaboration}\thanks{https://veritas.sao.arizona.edu}\\
        University of Utah\\
        E-mail: \email{payel.kar@utah.edu}}

\abstract{The TeV binary system \lsi has a compact object in an eccentric orbit around a Be star about 2 kpc from Earth. \lsi exhibits modulated gamma-ray emission around its 26.5 days orbit, mostly detectable at TeV energies around its apastron passage, with maximum flux during the $\phi = 0.55-0.65$ phase range. Multiple flaring episodes with nightly flux variability at TeV energies have been observed since its detection in 2006. At one time significant TeV emission was also detected in late 2010 from the source close to its periastron passage at superior conjunction. The TeV spectrum is well fitted by a power law with small variations of spectral index of $\sim2.6$ over the years. GeV, X-ray, and radio emission have been detected along the entire orbit, enabling detailed study of the modulation pattern and its super-orbital period, such comprehensive study of the \lsi orbit in the TeV regime has not been presented before. VERITAS has observed \lsi for over a decade now, accruing 200+ hours of data during different parts of its orbit. In this work, we have analyzed all available data for \lsi since September 2007 in individual phase bins of width $\Delta \phi = 0.1$ and performed a spectral analysis for two different parts of the orbit. TeV emission is now detected in 9 out of 10 phase bins, around the entire orbit for the first time in VERITAS data. Hint of spectral variation might also be present between different parts of the orbit. The implication of these results is discussed in the context of determining the nature of the unknown compact object (neutron star or microquasar) and a discussion of the absorption mechanisms in the system.}

\FullConference{35th International Cosmic Ray Conference --- ICRC2017\\
		10--20 July, 2017\\
		Bexco, Busan, Korea}

\begin{document}

\section{Introduction}
There are five high mass X-ray binary systems in our galaxy with majority of their power output at gamma ray energies ($>100$ GeV). All these binary systems hosts a massive O/B type star and a compact object (CO), out of which only PSR B1259-63 is confirmed to host a pulsar. For the rest i.e. \lsi, LS 5039, HESS J0632+057, and 1FGL J1018.6-5856, the nature of the CO is ambiguous, it could be either a neutron star or a blackhole, see \cite{2013A&ARv..21...64D} for review of these systems. Primarily, the orbital periods of gamma ray binaries modulate their flux across the electromagnetic spectrum but the exact physical mechanism producing the non thermal emissions in these systems is unclear. 

\lsi was first detected by VERITAS in 2007 and has been extensively observed since then \cite{2008ApJ...679.1427A,2015ICRC...34..818K,2016ApJ...817L...7A}. The flux from \lsi is strongly modulated from radio to TeV energies by the 26.5 day orbital period of the system \cite{2006A&A...459L..25B}. The system also exhibits superorbital flux modulations on a longer $\sim4.6$ year timescale which is attributed to the rapidly spinning Be star of $10-15 \ M_{\odot}$ in the system \cite{2016A&A...591A..76A}. There are two popular competing models for the very high energy (VHE) emission from \lsi based on the nature of the CO. The first model assumes the compact object to be a microquasar where the relativistic charged particles present in jets of the stellar blackhole produce the VHE emission, similar to active galactic nuclei \cite{2006A&A...459L..25B}. The second model assumes the systems hosts a neutron star; VHE emission is produced by accelerated charged particles at the shock front of the colliding winds from the neutron star and the massive Be star, similar to PSR B1259-63 \cite{2010MNRAS.403.1873Z}. While the microquasar model emulates three decades of radio observations and presents an impressive theory of astronomy of beats using periods associated with the system \cite{2016A&A...585A.123M}, no signature X-ray emission from an accretion disc has been observed from \lsi which would be essential to power its relativistic jets \cite{2010MNRAS.403.1873Z}. The neutron star model falls short in explaining the energy budget of the system, a much more powerful wind from the Be star would be required to overwhelm the typical pulsar wind of $\sim 10^{36} \ \mathrm{erg \ s^{-1}}$ expected in \lsi \cite{2007A&A...474...15R}. For detailed arguments for and against both the microquasar and neutron star models refer to \cite{2010MNRAS.403.1873Z,2007A&A...474...15R}. 

A magnetar-like event from the direction of \lsi observed by \textit{Swift} in 2012 sparked a debate of the first magnetar in a binary system \cite{2012ApJ...744..106T}. The highly magnetized neutron star would transition from propeller to ejector and back to propeller based on the surrounding matter densities along each eccentric orbital motion. We present the results from the long term monitoring of \lsi here and discuss the ejector-propeller flip-flop model in context. A similar decade long study of another gamma ray binary HESS J0632+057 and an outline of the binary discovery program by VERITAS is presented elsewhere in this conference \cite{2017ICRC..G}.

\section{VERITAS Observations and Data Analysis}
The VERITAS Observatory is an array of four Imaging Atmospheric Cherenkov Telescopes located at the Fred Lawrence Whipple Observatory (FLWO) in southern Arizona ($31^{\circ} 40'$ N, $110^{\circ} 57'$ W,  1.3km a.s.l.). The full array began operations in September 2007 (V4 configuration - Sep 2007 to July 2009) and has been upgraded twice. During the first upgrade, one of the telescopes was relocated, increasing the collection area and improving the angular resolution (V5 configuration September 2009 - July 2012). In the second upgrade, the PMTs and the telescope trigger system were upgraded, improving the the array's response at low energies (V6 configuration - Sep 2012 to current). In the current V6 configuration, VERITAS is sensitive to very high energy (VHE) gamma rays from 85 GeV to $>30$ TeV and detects a source with flux corresponding to $1\%$ of the Crab Nebula in less than 25 h. For details of the instrument and its performance see \cite{2015ICRC...34..771P}. 

\lsi is one of the most observed sources by VERITAS with over 240 hours of data on the source over the 10 years since its discovery at TeV energies. The orbital phase $\phi=0$ is set on \textbf{MJD 43366.275}, the date of discovery of \lsi as a variable radio source and the 26.5 d orbit is divided into equal 10 phases. The entire data set is distributed unevenly around the orbit with maximum exposure during phase range $\phi=0.5-0.8$ near the apastron passage. A summary of the VERITAS observations is shown in Table \ref{veritasobs}. 

\begin{table}[h]
	\centering
    \caption{VERITAS Decadal Dataset of \lsi Observations}%
    \label{veritasobs}
    \vspace{\baselineskip}
        \begin{tabular}{| c | c | c | c| }
           \hline\hline 
Observing & Instrument  & Quality Selected & Detection \\ Season & Epoch & Livetime [min] & Significance ($\sigma$)\\
		   \hline
 2007 / 2008 & V4 & 1518 & 6.2  \\ \hline
 2008 / 2009 & V4 & 2305 & 3.8  \\ \hline
 2009 / 2010 & V5 & 1207 & 0.7  \\ \hline
 2010 / 2011 & V5 &  933 & 4.6  \\ \hline
 2011 / 2012 & V5 & 1551 & 14.0 \\ \hline
 2012 / 2013 & V6 &  490 & 6.5  \\ \hline
 2013 / 2014 & V6 &  522 & 5.6  \\ \hline
 2014 / 2015 & V6 & 1746 & 21.4 \\ \hline
 2015 / 2016 & V6 & 1137 & 16.0 \\ \hline
 2016 / 2017 & V6 &  703 & 12.4 \\ \hline
 All & V4, V5, V6 & 12112& 29.2 \\ \hline
            \hline 
        \end{tabular}
\end{table}

Standard analysis for a point source is performed using advanced Boosted Decision Tree technique (BDT) \cite{2017APh....89....1K}. The source is significantly detected at energies $>300$ GeV in 7 out of 10 observing seasons. Spectral analysis is performed by dividing the orbit in two parts, the first part close to the apastron passage covering 3 phases bin $\phi=0.5\rightarrow0.8$ and the second for rest of the orbit covering phases $\phi=0.5\rightarrow0.8$. The later 7 phases are further divided into two parts based on the location of the compact object (CO) behind the Be star and its disc ($\phi=0.8\rightarrow0.2$) or in front of them ($\phi=0.2\rightarrow0.5$). Spectral analysis is also performed for both these phase intervals.

\section{Results}
\begin{figure}[t]
\centerline{\includegraphics[scale = 0.475]{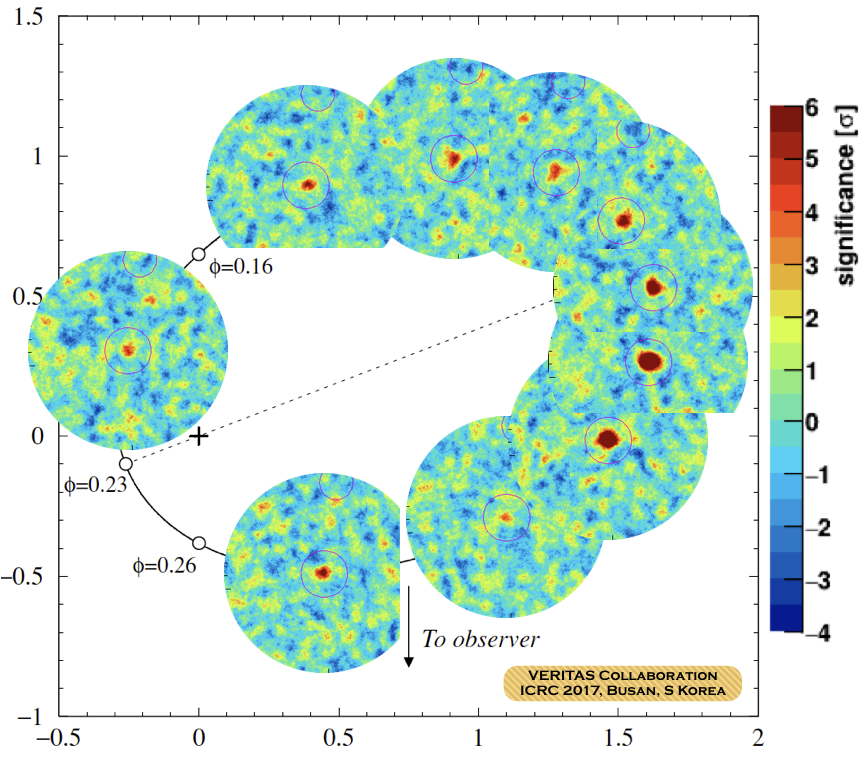}}
\caption{ VERITAS skymaps of \lsi arranged by orbital phase $\phi$. Image locations are approximate and illustrative only, the orbit is unresolved at TeV energies. The time progression is counter-clockwise. The orbital parameters used are from \cite{2005MNRAS.360.1105C}}%
\label{phaseskymap}
\end{figure}

\lsi is detected at $>5\sigma$ significance level in 9 out of 10 phase bins around the orbit. A summary of the results from each of the phase bin of width $\Delta \phi =0.1$ is summarized in Table \ref{phaseresulttable} and a shown in Fig \ref{phaseskymap}. In the single phase bin ($\phi=0.4-0.5$) the significance is $3.8\sigma$, 
\begin{table}[t]
	\centering
    \caption{\lsi Phase Binned Results}%
    \label{phaseresulttable}
    \vspace{\baselineskip}
        \begin{tabular}{| c | c | c | c | c | }
           \hline\hline
Phase bin & Pre-trial & Integral Flux above 300 GeV & \% of Crab	 & Exposure\\ ($\phi$) & Significance ($\sigma$) & [$\times10^{-12}$ \cms] & & [min]\\
		   \hline
$0.0 \rightarrow 0.1 $ & 5.6  & $ 2.90 \pm 0.61 $ & 2.8 & 992	\\ 
$0.1 \rightarrow 0.2 $ & 5.4  & $ 4.03 \pm 0.80 $ & 3.8 & 663	\\ 
$0.2 \rightarrow 0.3 $ & 5.1  & $ 2.77 \pm 0.60 $ & 2.6 & 1186	\\  
$0.3 \rightarrow 0.4 $ & 6.1  & $ 3.35 \pm 0.61 $ & 3.2 & 1178	\\ 
$0.4 \rightarrow 0.5 $ & 3.8  & $ <4.59 $ & $<4.3$ & 547	\\ 
$0.5 \rightarrow 0.6 $ & 15.9 & $ 6.57 \pm 0.54 $ & 6.2& 1289	\\ 
$0.6 \rightarrow 0.7 $ & 18.8 & $ 7.40 \pm 0.44 $ & 7 & 2517	\\ 
$0.7 \rightarrow 0.8 $ & 8.0  & $ 3.69 \pm 0.46 $ & 3.5 & 1630	\\
$0.8 \rightarrow 0.9 $ & 5.2  & $ 3.53 \pm 0.62 $ & 3.3 & 731	\\ 
$0.9 \rightarrow 0.0 $ & 5.0  & $ 3.81 \pm 0.85 $ & 3.6 & 543	\\ 
            \hline \hline
        \end{tabular}
\end{table} slightly below the defined significance level for claiming discovery. Integral flux estimate of the quiescent TeV emission from \lsi ($>300$ GeV) in phases $\phi=0.8\rightarrow0.4$ is $\sim3\%$ of the steady Crab Nebula flux. Spectral index of -2.6 is assumed to calculate the integral flux in each bin. 

\begin{figure}[b]
\centerline{\includegraphics[scale = 0.4]{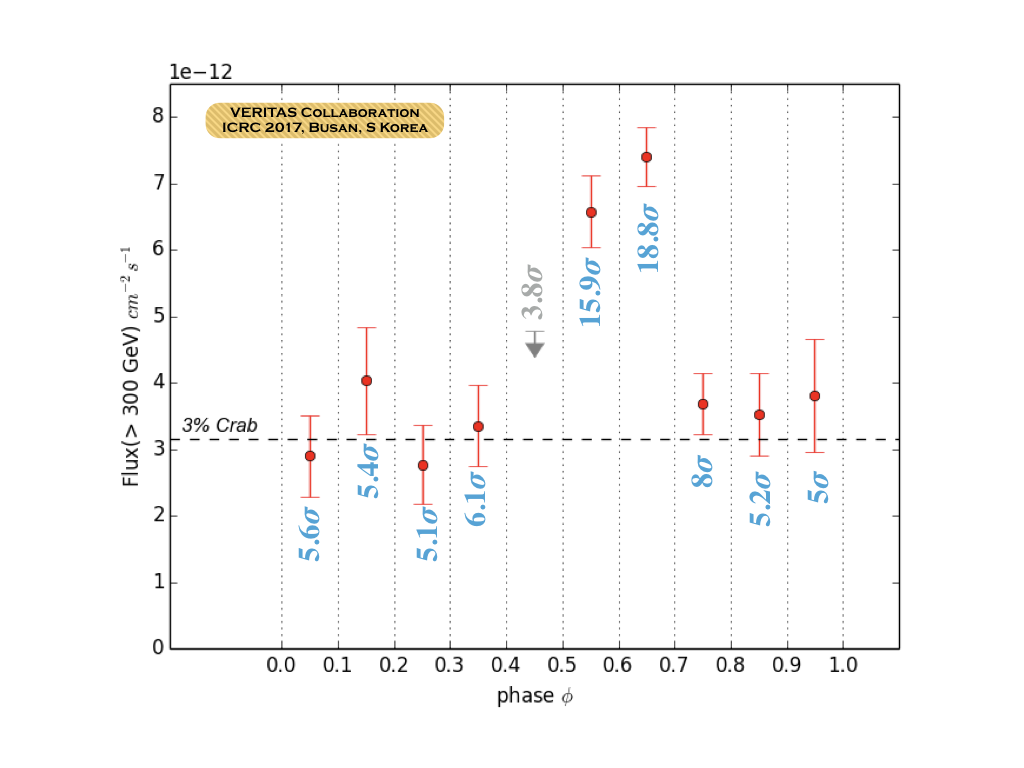}}
\caption{ Phase Binned flux of \lsi for decadal data. }%
\label{phaseflux}
\end{figure}

The spectrum \lsi is well fitted with a power-law of the form $$dN/dE = N_{0}\Bigg(\frac{E}{E_{0}}\Bigg)^{\Gamma}$$ A combined spectral analysis on the 3 phases near apastron ($\phi=0.5\rightarrow0.8$) yields a spectral index $\Gamma=-2.63\pm0.06_{stat}$ with reduced $\chi^{2}=1.1$, consistent with previously measured spectral indices calculated for historical TeV outbursts. For the rest of the orbit ($\phi=0.8\rightarrow0.5$) a spectral index $\Gamma=-2.81\pm0.16_{stat}$ is calculated with reduced $\chi^{2}=0.3$. A spectral study of the 2 subdivisions of the phase $\phi = 0.5\rightarrow0.8$ interval yields a spectral index $\Gamma=-2.86\pm0.21_{stat}$ when the CO is behind the disc of the Be star ($\phi=0.8\rightarrow0.2$) compared to $\Gamma=-2.62\pm0.22_{stat}$ when it is in front of the disc and Be star ($\phi=0.2\rightarrow0.5$). The errors on the spectral index are due to limited photon statistics and systematics.

\begin{figure}[t]
\centerline{\includegraphics[scale = 0.53]{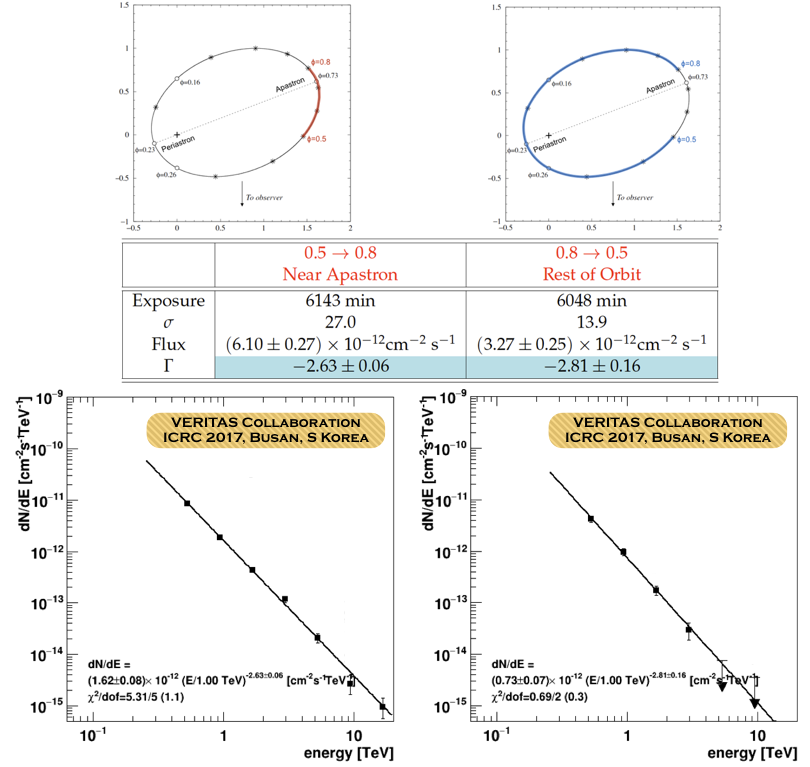}}
\caption{ Spectral energy distribution (SED) for \lsi for two parts of the orbit (parts of the orbit shown on top panels). SED on the \textit{left} is near apastron passage covering $\phi=0.5 \rightarrow 0.8$ and SED on the \textit{right} is for the rest of the orbit for $\phi=0.8 \rightarrow 0.5$. The orbital parameters shown on top panel are used from \cite{2005MNRAS.360.1105C} }%
\label{phasesed}
\end{figure}

\section{Discussion}
The nature of the CO of \lsi has been an active topic of debate since neither the microquasar model not the pulsar wind binary model are able to fully explain all existing observations. The baseline TeV emission and VHE outbursts near apastron could be explained by the neutron star flip flop model \cite{2012ApJ...744..106T}. This model assumes that \lsi hosts a young neutron star which flip flops between its ejector and propeller phases as the CO revolves around the Be star. Near periastron, the intense ram pressure of the decreted matter from the Be shrinks the Alv\'en radius of the neutron star. In this propeller state the neutron stars's magnetosphere extends beyond the co-rotation radius of the surrounding matter. The centrifugal force prevents the surrounding co-rotating matter from accreting onto the magnetic poles. The accumulated plasmon mass spirals around like a vortex with concomittant high voltages and shocks between the layers creating the quiescent TeV emmission that is observed \cite{2001cnoc.conf...50Z}. As the neutron star travels further from the dense stellar surroundings it transitions to from propeller phase to ejector phase, expelling the plasmon and producing TeV outbursts. The ejector-propeller flip-flop model fits the well with the flux variability in the various parts of the orbit as seen in Figure \ref{phaseflux}, a lower flux during propeller state of the neutron star (i.e. $\phi=0.8 \rightarrow 0.5$) and flaring TeV flux near apastron passage where the neutron star in its ejector expeling the accumulated matter from the previous state (i.e. $\phi=0.5 \rightarrow 0.8$). The orbit-to-orbit variability that is seen in the peak flux close to apastron passage \cite{2016ApJ...817L...7A} could be dependent on the mass of the accumulated matter during the propeller state of the neutron star. The microquasar model \cite{2016A&A...585A.123M} as well as the pulsar wind shock model \cite{2010MNRAS.403.1873Z} although both present reasonable explanations for the observed TeV outbursts near apastron, would need need adjustment to explain the newly discovered baseline TeV emission and flux variations for the rest of the orbit. 

The spectral energy distribution of VHE emission from \lsi close to its apastron in phases $\phi=0.5\rightarrow0.8$ has been stable over the years at $\sim2.6$. A similar spectral trend is seen for the phase range $\phi=0.2\rightarrow0.5$. In the phase range $\phi=0.8\rightarrow0.2$ hint of a softer index $\Gamma\simeq -2.86$ may be present but cannot be confirmed due to statistical uncertainty. Upper limits are calculated for the highest energy bins in the spectral energy distribution due to limited photon statistics and a cutoff above $\sim3$ TeV cannot be excluded. An evolving spectral behavior in different parts of the orbit could be a possibility for \lsi similar to LS 5039, where a cutoff at 6 TeV near its inferior conjunction passage is seen \cite{2015ICRC...34..885M}.  

The correlation seen between X-ray and TeV emission from \lsi favors a single zone model for the binary where charged particles produce synchrotron X-rays and VHE gamma rays by synchrotron self-Compton processes \cite{2015ICRC...34..818K}. A detailed multiyear correlation study at X-ray, GeV and TeV energies will will be presented in a forthcoming publication. Yet the theory of astronomical beats and the observed anti-correlation between X-ray luminosity and X-ray spectral index points towards a microquasar in the system. It is clear that \lsi is a much more complex source and further observations are required. Multiwavelegth observations specifically focussed on the rest of the orbit away from apastron could resolve the debate on the nature of the CO. Correlated emission patterns around different part of the orbits would be key to identifying the various particle populations at work. Fast variability studies at TeV energies on timescales of $\sim1000$ s by CTA will yield deeper insights on the size of the VHE emitting regions.  \\

\noindent\textbf{Acknowledgments}
This research is supported by grants from the U.S. Department of Energy Office of Science, the U.S. National Science Foundation and the Smithsonian Institution, and by NSERC in Canada. We acknowledge the excellent work of the technical support staff at the Fred Lawrence Whipple Observatory and at the collaborating institutions in the construction and operation of the instrument.

\bibliography{skeleton}

\providecommand{\href}[2]{#2}\begingroup\raggedright\begin{thebibliography}{10}

\bibitem{2013A&ARv..21...64D}
G.~{Dubus}, {\it {Gamma-ray binaries and related systems}},  {\em A\&A Rev.}
  {\bf 21} (Aug., 2013) 64, [\href{http://arxiv.org/abs/1307.7083}{{\tt
  arXiv:1307.7083}}].

\bibitem{2008ApJ...679.1427A}
V.~A. {Acciari} and {et al. (VERITAS Collaboration)}, {\it {VERITAS
  Observations of the {$\gamma$}-Ray Binary LS I +61 303}},  {\em ApJ} {\bf
  679} (June, 2008) 1427--1432, [\href{http://arxiv.org/abs/0802.2363}{{\tt
  arXiv:0802.2363}}].

\bibitem{2015ICRC...34..818K}
P.~{Kar} and {VERITAS Collaboration}, {\it {Long-term VERITAS monitoring of LS
  I 61 +303 in conjunction with X-ray, and GeV observation campaigns}},  in
  {\em 34th International Cosmic Ray Conference (ICRC2015)}, vol.~34 of {\em
  International Cosmic Ray Conference}, p.~818, July, 2015.
\newblock \href{http://arxiv.org/abs/1508.0667}{{\tt arXiv:1508.0667}}.

\bibitem{2016ApJ...817L...7A}
S.~{Archambault} and {et al. (VERITAS Collaboration)}, {\it {Exceptionally
  Bright TeV Flares from the Binary LS I +61 303}},  {\em ApJL} {\bf 817}
  (Jan., 2016) L7, [\href{http://arxiv.org/abs/1601.0181}{{\tt
  arXiv:1601.0181}}].

\bibitem{2006A&A...459L..25B}
V.~{Bosch-Ramon}, J.~M. {Paredes}, G.~E. {Romero}, and M.~{Rib{\'o}}, {\it {The
  radio to TeV orbital variability of the microquasar LS I +61 303}},  {\em
  A\&A} {\bf 459} (Nov., 2006) L25--L28.

\bibitem{2016A&A...591A..76A}
M.~L. {Ahnen} and {et al. (MAGIC Collaboration)}, {\it {Super-orbital
  variability of LS I +61{ }303 at TeV energies}},  {\em A\&A} {\bf 591} (June,
  2016) A76, [\href{http://arxiv.org/abs/1603.0697}{{\tt arXiv:1603.0697}}].

\bibitem{2010MNRAS.403.1873Z}
A.~A. {Zdziarski}, A.~{Neronov}, and M.~{Chernyakova}, {\it {A compact pulsar
  wind nebula model of the {$\gamma$}-ray-loud binary LS I +61 303}},
  {\em MNRAS} {\bf 403} (Apr., 2010) 1873--1886,
  [\href{http://arxiv.org/abs/0802.1174}{{\tt arXiv:0802.1174}}].

\bibitem{2016A&A...585A.123M}
M.~{Massi} and G.~{Torricelli-Ciamponi}, {\it {Origin of the long-term
  modulation of radio emission of LS I +61 303}},  {\em A\&A} {\bf 585} (Jan.,
  2016) A123, [\href{http://arxiv.org/abs/1511.0562}{{\tt arXiv:1511.0562}}].

\bibitem{2007A&A...474...15R}
G.~E. {Romero}, A.~T. {Okazaki}, M.~{Orellana}, and S.~P. {Owocki}, {\it
  {Accretion vs. colliding wind models for the gamma-ray binary LS I +61 303:
  an assessment}},  {\em A\&A} {\bf 474} (Oct., 2007) 15--22,
  [\href{http://arxiv.org/abs/0706.1320}{{\tt arXiv:0706.1320}}].

\bibitem{2012ApJ...744..106T}
D.~F. {Torres}, N.~{Rea}, P.~{Esposito}, J.~{Li}, Y.~{Chen}, and S.~{Zhang},
  {\it {A Magnetar-like Event from LS I +61 303 and Its Nature as a Gamma-Ray
  Binary}},  {\em ApJ} {\bf 744} (Jan., 2012) 106,
  [\href{http://arxiv.org/abs/1109.5008}{{\tt arXiv:1109.5008}}].

\bibitem{2017ICRC..G}
G.~{Maier} and {VERITAS Collaboration}, {\it {VHE Observations of Galactic
  binary systems with VERITAS}},  in {\em these proceedings}.

\bibitem{2015ICRC...34..771P}
N.~{Park} and {VERITAS Collaboration}, {\it {Performance of the VERITAS
  experiment}},  in {\em 34th International Cosmic Ray Conference (ICRC2015)},
  vol.~34 of {\em International Cosmic Ray Conference}, p.~771, July, 2015.

\bibitem{2017APh....89....1K}
M.~{Krause}, E.~{Pueschel}, and G.~{Maier}, {\it {Improved {$\gamma$}/hadron
  separation for the detection of faint {$\gamma$}-ray sources using boosted
  decision trees}},  {\em Astroparticle Physics} {\bf 89} (Mar., 2017) 1--9,
  [\href{http://arxiv.org/abs/1701.0692}{{\tt arXiv:1701.0692}}].

\bibitem{2005MNRAS.360.1105C}
J.~{Casares}, I.~{Ribas}, J.~M. {Paredes}, J.~{Mart{\'{\i}}}, and C.~{Allende
  Prieto}, {\it {Orbital parameters of the microquasar LS I +61 303}},  {\em
  MNRAS} {\bf 360} (July, 2005) 1105--1109,
  [\href{http://arxiv.org/abs/astro-ph/0504332}{{\tt astro-ph/0504332}}].

\bibitem{2001cnoc.conf...50Z}
R.~{Zamanov}, J.~{Marti}, and P.~{Marziani}, {\it {Be/X-ray Binary LSI+61303 in
  Terms of Ejector-Propeller Model}},  in {\em The Second National Conference
  on Astrophysics of Compact Objects}, p.~50, Sept., 2001.
\newblock \href{http://arxiv.org/abs/astro-ph/0110114}{{\tt astro-ph/0110114}}.

\bibitem{2015ICRC...34..885M}
C.~{Mariaud}, P.~{Bordas}, F.~{Aharonian}, G.~{Dubus}, M.~{B{\"o}ttcher},
  M.~{de Naurois}, C.~{Romoli}, and V.~{Zabalza}, {\it {H.E.S.S. observations
  of LS 5039}},  in {\em 34th International Cosmic Ray Conference (ICRC2015)},
  vol.~34 of {\em International Cosmic Ray Conference}, p.~885, July, 2015.
\newblock \href{http://arxiv.org/abs/1509.0579}{{\tt arXiv:1509.0579}}.

\end{thebibliography}\endgroup
% \begin{thebibliography}{99}
% \bibitem{...}
% ....
%
% \end{thebibliography}

\end{document}